\documentclass[preprint,12pt]{elsarticle}

\usepackage{geometry}[margins=1in]
\usepackage{amssymb}
\usepackage{amsmath}
\usepackage{float}
\usepackage{xcolor}
\usepackage{url} 

\journal{Coastal Engineering}

\begin{document}

\begin{frontmatter}

\title{Global Location-Invariant Peak Storm Surge Prediction} 

\author[inst1]{Benjamin Pachev\corref{cor1}}
\ead{bpachev@umass.edu} 
\affiliation[inst1]{
  institution={College of Information and Computer Sciences, University of Massachusetts Amherst},
  city={Amherst},
  postcode={01002},
  state={Massachusetts},
  country={USA},
}

\author[inst5]{Prateek Arora}
\ead{arorap@berkeley.edu}
\affiliation[inst5]{
  institution={Department of Civil and Environmental Engineering, University of California, Berkeley},
  city={Berkeley},
  postcode={94704},
  state={California},
  country={USA},
}

\author[inst2]{Jinpai Zhao}
\ead{max.zhao@utexas.edu} 
\affiliation[inst2]{
  institution={Oden Institute for Computational Engineering and Sciences, The University of Texas at Austin},
  city={Austin},
  postcode={78712},
  state={Texas},
  country={USA},
}

\author[inst3,inst4]{Eirik Valseth}
\ead{eirik.valseth@nmbu.no,eirikva@simula.no} 
\affiliation[inst3]{organization={ Department of Mechanical Engineering and Technology Management, Norwegian University of Life Sciences},
            addressline={Drøbaksveien 31}, 
            city={Ås},
            postcode={1430}, 
            country={Norway}}
\affiliation[inst4]{organization={Department of Numerical Analysis and Scientific Computing, Simula Research Laboratory},
            addressline={Kristian Augusts gate 23}, 
            city={Oslo},
            postcode={0164}, 
            country={Norway}}

 \cortext[cor1]{Corresponding author}

\begin{abstract}
Storm surge is a significant threat to coastal communities across the globe, responsible for loss of life and enormous property damage. Consequently, significant efforts have been expended to develop high-fidelity physics-based models for storm surge prediction. However, such models are often extremely computationally expensive and require supercomputing resources. In recent years, there has been a growing trend towards data-driven surrogate models, which approximate the capabilities of high-fidelity models at a tiny fraction of the computational cost.

Most datasets of high-fidelity storm surge model output are limited to narrow geographical regions, with the majority focused on the continental United States and China. This trend is reflected in the scope of existing storm surge surrogate models. In this work, we present a novel dataset for training storm surge surrogate models with global applicability. The dataset consists of high-resolution peak surge output from the ADvanced CIRCulation (ADCIRC) model for over 15,000 landfalling synthetic storms distributed across the world. To the author's knowledge, it is the largest dataset of its kind ever assembled, and is unique in its global scope.

We additionally present a machine learning model for peak storm surge based on computer vision architecture. The model is trained on our new global dataset and can accurately predict maximum storm surge in disparate geographical regions - including those for which few or no surrogate models exist. Both the dataset and accompanying model are publicly available, with the aim to support the development of additional storm surge models with global reach.

\end{abstract}


\end{frontmatter}

\section{Introduction}
\label{sec:introduction}

Storm surge is a major natural hazard, responsible for significant loss of life \cite{rappaport2014fatalities} and billions of dollars in economic damage annually \cite{noaabillion}. Accurate modeling of storm surge is important for emergency management and development of resilient coastal infrastructure. Due to the high computational costs associated with traditional physics-based models of storm surge, efforts in recent years have focused on inexpensive surrogate and data-driven models, see e.g.,~\cite{LEE2006483,LEE200863,LEE20091200, ian2022performance, ALKAJBAF2020106184, nhess-12-3799-2012, kim2015time, hashemi2016efficient, DBLP:journals/corr/BezuglovBS16, kim2018surrogate, LEE2021104024,pachev2023framework}. Here we give a brief overview of some representative and recent works.

The   framework developed  by Jia and Taflanidis in 2013~\cite{jia2013kriging}, later expanded in 2016, established the foundational methodology for modern storm surge emulation by coupling Kriging (Gaussian Process Metamodeling, or GPM) with Principal Component Analysis (PCA) \cite{jia2013kriging,jia2016surrogate}. This GPM+PCA approach was the first to solve the high-dimensional output problem, enabling the rapid approximation of full-field surge responses from a small database of high-fidelity simulations \cite{jia2016surrogate}. Research following this foundational work has focused on operationalizing the framework by solving key implementation challenges. Recent work includes the development of sophisticated dual-surrogate systems--using a primary regression model for surge magnitude and a secondary classification model to handle the binary wet/dry state of inland nodes \cite{kyprioti2022integration, kyprioti2021improvements} and advanced imputation strategies to manage the sparse, incomplete data common in expensive simulation databases \cite{kyprioti2021improvements}.

In parallel to the GPM-based advancements, a separate track leveraging Artificial Neural Networks (ANNs) and Deep Learning (DL) architectures like Long Short-Term Memory (LSTM) and Convolutional Neural Networks (CNNs) has also matured, proving highly effective at modeling non-linear surge dynamics, especially when trained on large observational or reanalysis datasets \cite{ hashemi2016efficient, tiggeloven2021exploring}. A significant trend has been the convergence of these two divergent paths. For example, recent surrogate models like C1PKNet explicitly adopt the PCA dimensionality reduction technique from the GPM framework but replace the Kriging engine with a more powerful 1D-Convolutional Neural Network (CNN) \cite{lee2021rapid}. This hybrid-DL approach merges the structural advantages of the original flexible framework with the pattern-recognition capabilities of deep learning.

In our 2023 paper~\cite{pachev2023framework}, we introduced a  novel surrogate model for peak storm surge prediction based on a multi-stage approach. In addition, the formulation allowed independent surge predictions at each point in the domains, i.e., introducing flexibility to make predictions for locations not present in the training data set. 
In the present work, our focus remains on peak storm surge prediction, which is more amenable to prediction by purely data-driven models. Additionally, the maximum surge is a key quantity of interest in applications such as flood risk assessment and emergency management \cite{dietrich2013real}. In such applications, thousands or tens of thousands of surge model evaluations are used to assess inundation risks \cite{yang2019objective}. Only the maximum surge profile is needed, not the entire timeseries, which provides the motivation for inexpensive surrogate models of peak surge \cite{gharehtoragh2024evolving}.

In this work we address two primary limitations of data-driven surrogate models: data scarcity and regional specificity. While the small size of training datasets is a significant challenge for surrogate modeling, it is still possible to obtain good predictive accuracy \cite{kyprioti2021improvements}. However, this accuracy comes at the price of concentrating a high number of storms in a small geographic area \cite{ayyad2022machine}. Consequently, most storm surge models are inapplicable outside of their training region. This is an issue of model formulation as well as a limitation in the training data. All surrogate models are maps from an input space to an output space. Typically, the output space has fixed dimension and corresponds to the storm surge at a concrete set of points. This implicitly encodes the geographic region of interest into the model architecture. While the traditional region-centric approach can perform well in the area of interest, it scales poorly. For each new area, an entirely new training dataset and model are required. Given the computational cost of creating training data, this undercuts the entire premise of surrogate modeling, which is to provide a cheap alternative to high-fidelity models.

Here we present a novel database of storm surge simulations for over 15,000 synthetic storms on a high-resolution global mesh. The database is an order of magnitude larger than existing datasets - most of which contain fewer than 1,000 storms~\cite{johnson2023coastal,designsafe2021storms,gharehtoragh2024evolving}. As a numerical model, we use the ADvanced CIRculation (ADCIRC) model~\cite{luettich1992adcirc,pringle2021global} with inputs from established operational simulation frameworks.
The storms are distributed across the globe and provide significant variation in bathymetric and coastline features at the point of landfall.

The remainder of this work is organized as follows: Section \ref{sec:data} describes the novel simulation database, Section \ref{sec:methodology} details the feature engineering and surrogate model architecture, Section \ref{sec:results} analyzes the performance of our surrogate model and finally, Section \ref{sec:conclusion} summarizes our results and discusses some potential directions for future research.

\section{Data}
\label{sec:data}

The greatest limiting factor in developing surrogate models is the training data \cite{kyprioti2021improvements}. In the case of storm surge, observations of real events are typically restricted to a highly limited set of locations. While extremely useful for validating predictions, observational data is unsuitable for training high-resolution machine learning models of storm surge. To overcome the spatial paucity of observational data, high-resolution output from physics-based models is used instead of actual observations. As the number of historical hurricanes with meaningful storm surge is limited, \textit{synthetic storms} are used instead.

\subsection{Meteorological Forcing}

Physics-based numerical models of storm surge based on the two-dimensional Shallow Water Equations, such as ADCIRC, take wind and pressure fields as their primary input, and output water elevations. These meteorological inputs can come from high-resolution meteorological models such as the Global Forecast System (GFS) \cite{GFS}. High-resolution meteorological data is expensive and difficult to generate for synthetic storms, so we instead use the parametric Holland model \cite{holland1980analytic}. Parametric wind models use the projected storm track, radius to maximum winds, central pressure, and other values to generate wind and pressure fields.

In order to generate synthetic training data, we need a source of synthetic meteorological inputs which accurately reflects the distribution of real hurricanes, but contains enough storms to train an accurate model. In this work, we use the STORM IBTrACS dataset~\cite{bloemendaal2022storm, bloemendaal2020generation}, which contains 10,000 years of simulated tropical cyclones under present conditions. In total, there are well over a hundred thousand synthetic hurricanes in the dataset. Storm track parameters are provided every three hours and consist of eye latitude, eye longitude, radius to maximum winds, minimum pressure, and maximum wind velocity. Storms are subdivided into six ocean basins, which are listed in Table \ref{tab:basins}.

\begin{table}[H]
    \centering
    \begin{tabular}{|c|c|}
        \hline
        Basin &
        Name \\ \hline 
        NA & North Atlantic \\ \hline
        EP & East Pacific \\ \hline
        NI & North Indian \\ \hline
        SI & South Indian\\ \hline
        WP & West Pacific \\ \hline
        SP & South Pacific \\ \hline
    \end{tabular}
    \caption{Abbreviations and full names for the six ocean basins in the IBTrACS dataset.}
    \label{tab:basins}
\end{table}

\subsection{ADCIRC}

We use ADCIRC for generating high-resolution storm surge data with meteorological inputs taken from the STORM IBTrACS dataset. The raw track data was converted into ADCIRC's symmetric Holland model format. The same global mesh was used for all runs, which consists of 12.8 million nodes and 24.9 million triangular elements and is used operationally in NOAA's STOFS-2D-Global ocean circulation model \cite{Seroka2023-mm}.

Due to computational cost, it is infeasible to do ADCIRC runs for the entire input set. Consequently, we focus on landfalling storms of moderate to severe intensity. Filtering out low-intensity and non-landfalling storms leaves over 15,000 tropical cyclones. This still represents an extremely large number of ADCIRC runs and high computational cost.

\subsection{Storm Packing}

To further reduce the computational cost, we pack multiple cyclones into each ADCIRC run. While ADCIRC allows for multiple storms for some input types, it did not support multiple inputs for the symmetric Holland model. Consequently, we modified the ADCIRC codebase to add this feature. A maximum of one cyclone per ocean basin was used to prevent interference between storms. Extensive testing comparing single-storm runs to multi-storm runs revealed that the peak storm surge generated by each cyclone was unaffected by the presence of additional cyclones in other ocean basins. 

Using the packing scheme reduces the total number of needed runs from over 15,000 to only 3,000. Performance benchmarking showed no significant increase in runtime for the multi-storm simulations compared to those with only one storm. The packing scheme ultimately reduced costs by over a factor of five. The lack of performance difference between multi-storm and single-storm runs is not surprising, because the only difference is in the generation of the meteorological forcing, which is not the dominant cost in ADCIRC. We note that a similar speedup could likely be achieved by using multiple smaller regional meshes. However, this would require significant effort, and could negatively impact accuracy.

\subsection{Tidal Spin Ups}

Realistic initial conditions are important for accurate ADCIRC simulations. Initial conditions are typically generated by running ADCIRC without meteorological forcing enabled. This is referred to as a \textit{tidal spin up}, as only tidal forcing is used for the run. We performed 120 seven day tidal spin up runs to create a set of realistic initial conditions for the storm simulations. Astronomical tidal forces in ADCIRC are ultimately a function of the date and time at which the simulation starts. Starting times for the spin up runs were selected to sample a wide variety of tidal conditions. A full list is included in the supplementary material. The number of unique tidal spin ups was limited by computational constraints, but is sufficient to minimize duplication of initial conditions in the dataset. The primary purpose of varying the tidal forcing and initial conditions was to avoid biasing the model towards a specific set of initial conditions.

\subsection{Computational Setup}

All simulations were performed on the Frontera system of the Texas Advanced Computing Center (TACC)~\cite{stanzione2020frontera}. Each ADCIRC run used 500 CPU cores and took an average of 1.48 hours to complete. The total computational cost across all 3,000 runs was 2.22 million CPU core-hours.

\subsection{Data Availability}

Our dataset is publicly available on the DesignSafe platform \cite{designsafe-adcirc}. 

\section{Surrogate Modeling Methodology}
\label{sec:methodology}

Most existing data-driven models predict storm surge at a fixed set of points in a predefined geographical region. This simplifies the prediction problem, but limits model applicability to the region of interest. Our aim is to create a generally applicable model. This is reflected in the dataset we have created, which enjoys global coverage. In addition to a global dataset, we employ a modeling framework that is not restricted to a predetermined set of output locations.

\subsection{Feature Extraction}

The dataset of ADCIRC simulations consists of maximum water elevations on an irregular mesh. For each storm, we interpolate the irregular maximum water elevation onto a regular lat-lon grid centered on the storm landfall location. The bathymetry is also interpolated onto the grid. The meteorological forcing fields are computed at grid points from the storm track data using the parametric Holland model. The wind and pressure are evaluated over a fixed temporal window centered at the time of landfall. The extent and resolution for the prediction grid and temporal window for forcing data are configurable. Changing the prediction grid changes the learning problem. A larger spatial window about the landfall allows for prediction of the full impact of a storm, but also makes it harder to zero in on the most extreme surges, which tend to occur near landfall. Increasing the resolution enables more detailed predictions, but increases the cost to train the model. Furthermore, increased resolution is unhelpful if it exceeds the resolution of the underlying mesh.

We selected a window size of 2.5 by 2.5 degrees, with a resolution of 128 by 128. A power of two is used to simplify the use of vision model architectures. Forcing data is evaluated at three-hour intervals starting 24 hours before landfall and ending 12 hours after landfall, for a total of 13 samples. This results in a total of 39 input channels for meteorological forcing (13 each for pressure and two wind components). One is used for bathymetry. We add another masked feature representing the presence or absence of land. This is necessary because the input mesh does not cover the interior major landmasses - only the ocean and coastal regions. 

\begin{figure}[H]
    \centering
    \includegraphics[width=1.0\linewidth]{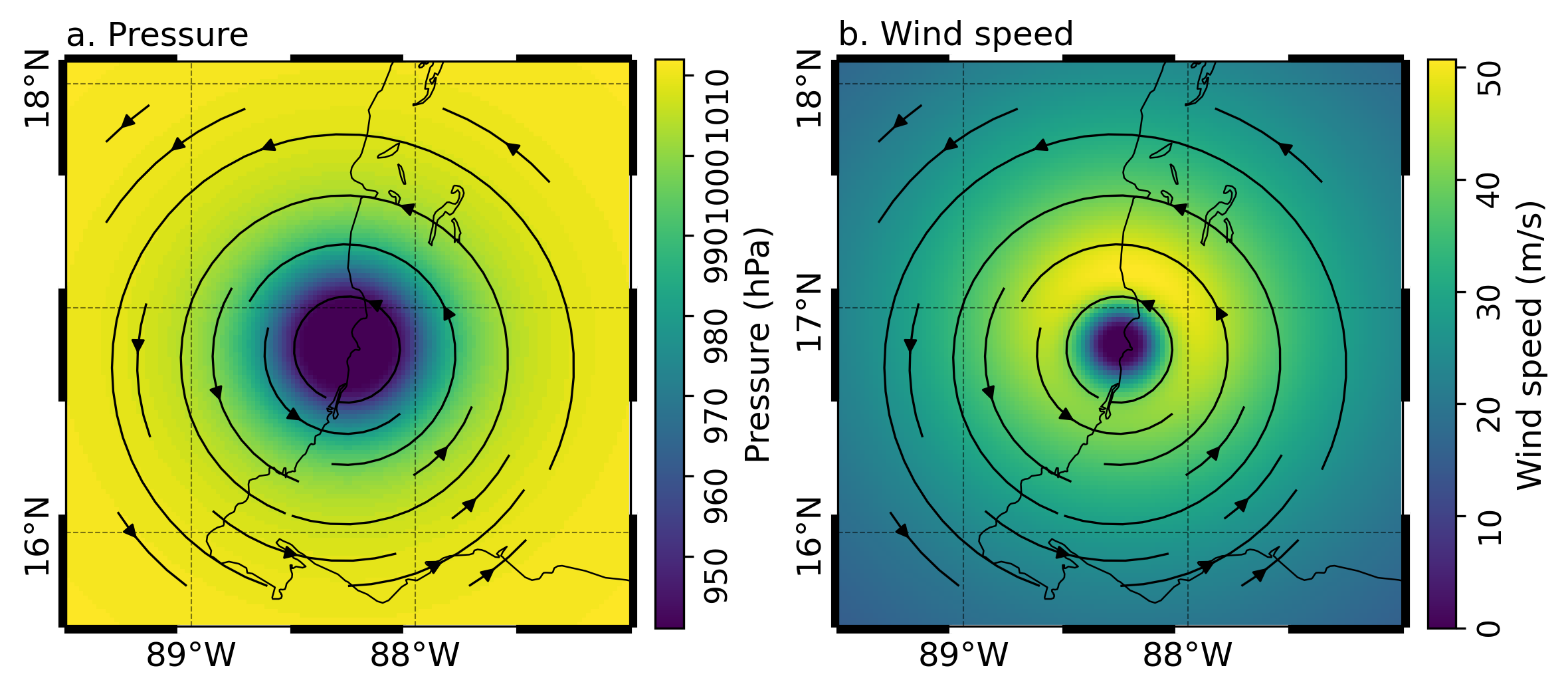}
    \caption{Pressure (mbar) and wind magnitude (m/s) at landfall for a sample synthetic storm. Wind direction is indicated with arrows, and the coastline is imposed on the plot.}
    \label{fig:pressure_winds}
\end{figure}

In total, there are 41 input features defined on the regular 128 by 128 grid. This setup is amenable to using computer vision architectures and requires little feature engineering other than interpolation onto a regular grid. Figure \ref{fig:pressure_winds} illustrates the gridded pressure and wind fields at the time of landfall for a sample synthetic storm.

\subsection{Model Architecture}

We employ a UNet architecture for predicting peak storm surge from the gridded input features. The UNet is a fully convolutional encoder-decoder network with skip connections, originally developed for biomedical image segmentation but widely adopted for dense prediction tasks involving structured spatial outputs \cite{huang2020unet}.

Let $\mathbf{X} \in \mathbb{R}^{C \times H \times W}$ denote the input tensor, where $C = 41$ is the number of input channels and $H = W = 128$ is the spatial resolution. The input channels comprise 13 temporal snapshots each of sea-level pressure, wind velocity (two components), along with bathymetry and a binary land mask. The network learns a mapping $f_\theta: \mathbb{R}^{C \times H \times W} \rightarrow \mathbb{R}^{H \times W}$ parameterized by weights $\theta$, producing the predicted maximum water elevation field $\hat{\zeta} = f_\theta(\mathbf{X})$.

The encoder pathway extracts hierarchical features through five successive stages. Each stage applies two $3 \times 3$ convolutions with ReLU activations, followed by $2 \times 2$ max pooling. The channel dimension doubles at each stage while spatial dimensions are halved, with base width $C_0 = 64$. 

The decoder pathway reconstructs the full spatial resolution through five corresponding stages. At each stage, a $2 \times 2$ transposed convolution upsamples the features, which are then concatenated with the encoder features at matching resolution via skip connections, followed by two $3 \times 3$ convolutions with ReLU activations. These skip connections preserve fine-grained spatial information that would otherwise be lost during encoding, enabling precise localization of surge maxima near complex coastline geometries.

A final $1 \times 1$ convolution projects the output to a single channel, yielding the predicted surge field $\hat{\zeta} \in \mathbb{R}^{H \times W}$. 
The model is trained by minimizing the mean squared error between predictions and ADCIRC-simulated peak surge:
\begin{equation}
    \mathcal{L}(\theta) = \frac{1}{NHW} \sum_{i=1}^{N} \sum_{h=1}^{H} \sum_{w=1}^{W} \left( \hat{\zeta}_i(h,w) - \zeta_i(h,w) \right)^2,
\end{equation}
where $N$ is the number of training samples.
We use the AdamW optimizer \cite{loshchilov2018decoupled} with learning rate $10^{-6}$ and weight decay $10^{-5}$, cosine annealing of the learning rate, and gradient clipping with maximum norm $1.0$.

\subsection{Data Partitioning}

\begin{figure}
    \centering
    \includegraphics[width=1.0\linewidth]{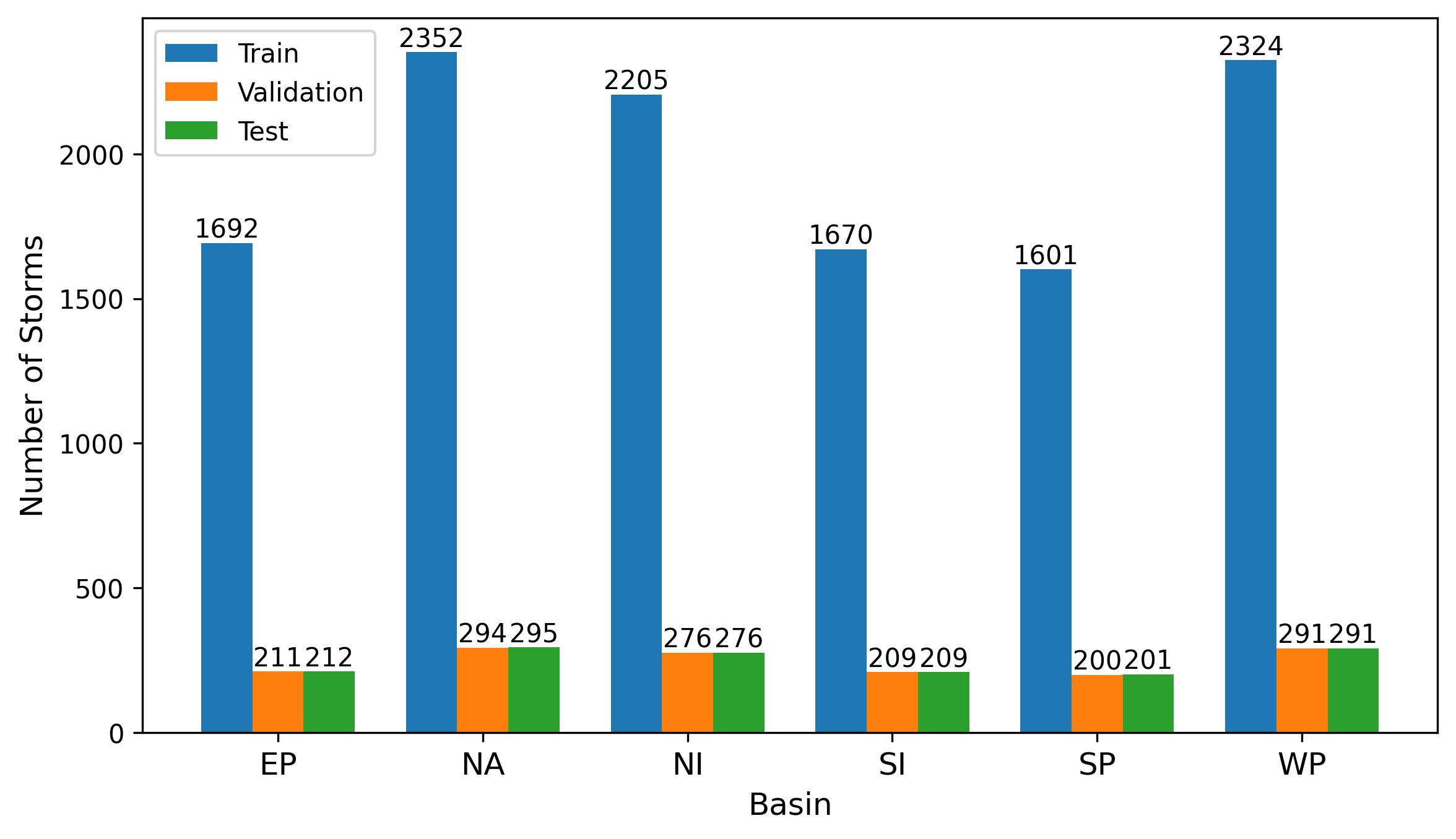}
    \caption{Caption}
    \label{fig:data_distribution}
\end{figure}

We randomly partitioned the data into a training set containing 80\% of the storms, a validation set with 10\%, and a holdout test set with 10\%. Figure \ref{fig:data_distribution} reports the split by ocean basin.

\subsection{Other Architectures}

In addition to UNet, we implemented other vision architectures such as basic convolutional neural networks (CNN), SegNet for regression \cite{badrinarayanan2017segnet}
and a simple feed-forward network. While UNet had the best performance by far, we include results from the other architectures for benchmarking and illustrative purposes. 

\subsection{Computational Resources}

All model training used the Vista system of the Texas Advanced Computing Center. Models were trained in parallel using up to four GH200 GPUs. Training times ranged from thirty minutes to a few hours, depending on model and dataset size.

\subsection{Code and Model Availability}

The code used to generate the features and train the models is publicly available at https://github.com/UT-CHG/adcirc-rom. Trained models, input datasets, and code needed to evaluate the models are available on DesignSafe \cite{designsafe-ml}.

\section{Results}
\label{sec:results}

\begin{figure}[H]
    \centering
    \includegraphics[width=1.0\linewidth]{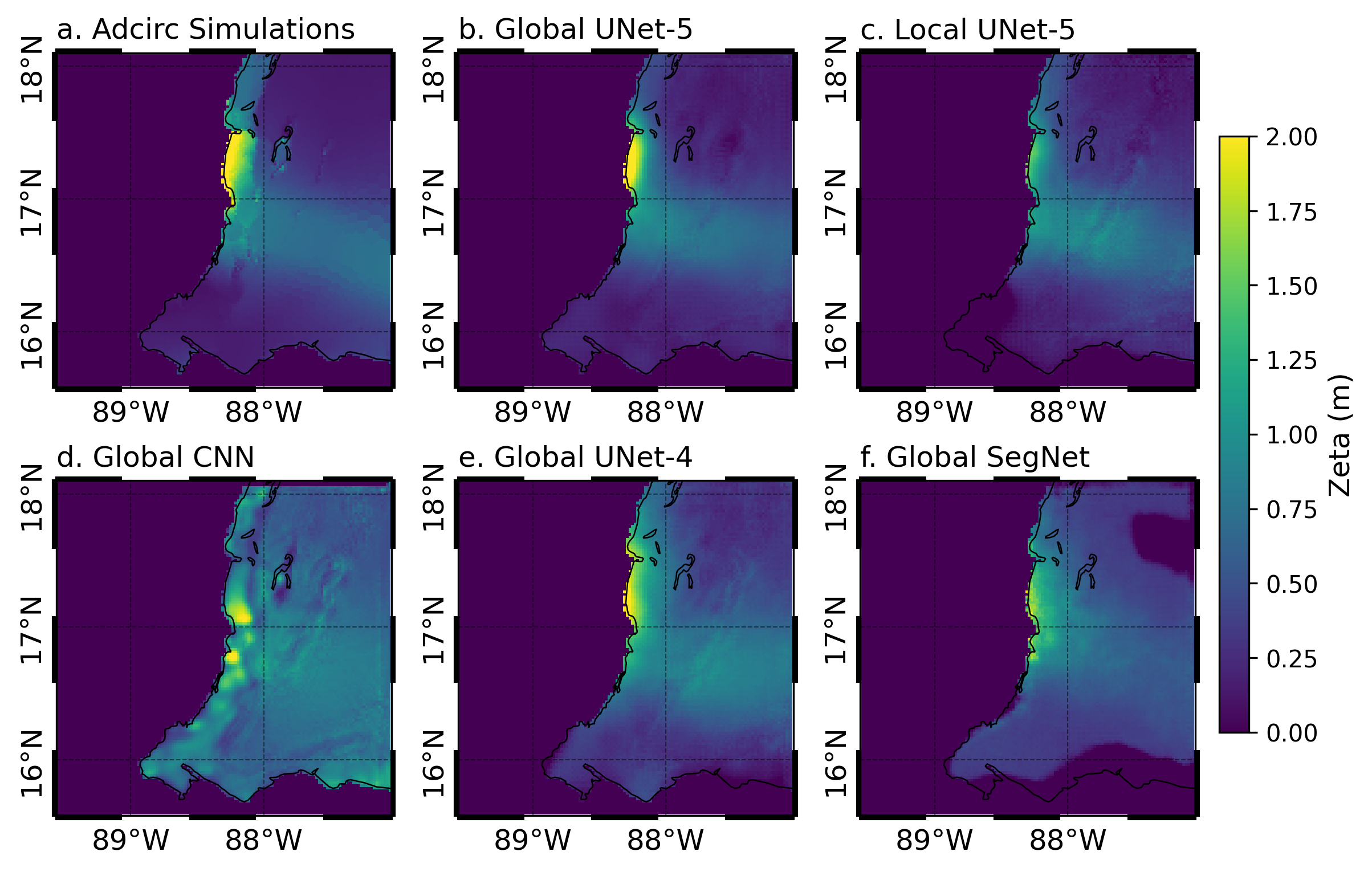}
    \caption{Predictions of maximum water levels for each model architecture for a sample storm, compared to the ADCIRC training data.}
    \label{fig:model_comparison}
\end{figure}

The best performing vision architecture was UNet with 5 encoder/decoder stages, which we will refer to as UNet-5. 
We report RMSE for different vision architectures in Table \ref{tab:different_archs}.
The U-Net architecture captures global contextual information through the encoder layers, while local spatial features are preserved through skip connections. 
The skip connections transfer high-resolution feature maps from the encoder to the corresponding upsampled decoder layers. 
This combination of global and local information enables the network to capture the spatial patterns of pressure and wind fields, see e.g., Figure \ref{fig:pressure_winds}, which become more pronounced as the storm approaches the coastline. Figure \ref{fig:model_comparison} depicts the differences in model predictions among architectures for a sample storm. The 'Local UNet-5' model uses the UNet-5 architecture but is trained only on data from the same ocean basin as the sample storm.

\begin{table}[H]
    \centering
    \begin{tabular}{|c|p{2.8cm}|p{2.8cm}|}
        \hline
        Model &
        RMSE (near land) &
        RMSE (all points) \\ \hline 
        
        CNN & 1.38 & 0.69\\ \hline
        Unet-4 & 0.66 & 0.38\\ \hline
        \textbf{UNet-5} & \textbf{0.51} & \textbf{0.33}\\ \hline
        SegNet & 0.69 & 0.44\\ \hline
    \end{tabular}
    \caption{RMSE (meters) for different architectures on the same holdout test data.}
    \label{tab:different_archs}
\end{table}

\subsection{Local vs Global}

To assess the impact of regionality, we trained separate (local) UNet-5 models for each basin. We compared the performance of the local models to the global UNet-5 model, and have summarized the results in Table~\ref{tab:local_vs_global}. In addition to the RMSE across the full prediction window, we also report the RMSE in coastal regions using a three-pixel-dilation window centered on land.

The global model outperforms basin-specific local models in terms of RMSE across all basins, achieving error reductions of approximately 10\% in the NA basin and 28\% in the NI basin. The availability of a larger and more heterogeneous training dataset enables the global model to generalize storm-surge behavior near complex coastal geometries that may be underrepresented in basin-specific training. The significance of this result is that inclusion of training data from entirely separate geographical regions substantially improves model performance. This suggests that the model is in fact learning storm surge mechanics instead of memorizing the surge response profile of a particular section of the coastline. Appendix A contains a set of plots that compare the local and global UNet-5 model predictions for multiple storms in each ocean basin.

\begin{table}[H]
    \centering
    \begin{tabular}{|c|p{2.8cm}|p{2.8cm}|p{2.8cm}|p{2.8cm}|}
        \hline
        Basin &
        Local Model (near coastline) &
        Local Model (all points) &
        Global Model (near coastline) &
        Global Model (all points) \\ \hline 
        
        NA & 0.73 & 0.38 & 0.66 & 0.34 \\ \hline
        EP & 0.26 & 0.19 & 0.23 & 0.18 \\ \hline
        NI & 0.99 & 0.48 & 0.71 & 0.40 \\ \hline
        SI & 0.74 & 0.49 & 0.64 & 0.48 \\ \hline
        WP & 0.34 & 0.27 & 0.34 & 0.27 \\ \hline
        SP & 0.49 & 0.31 & 0.38 & 0.28 \\ \hline
    \end{tabular}
    \caption{RMSE (meters) in different basins for local and global models on the same holdout test data.}
    \label{tab:local_vs_global}
\end{table}


\subsection{Historical Storms}

\begin{figure}[H]
    \centering
    \includegraphics[width=1.0\textwidth]{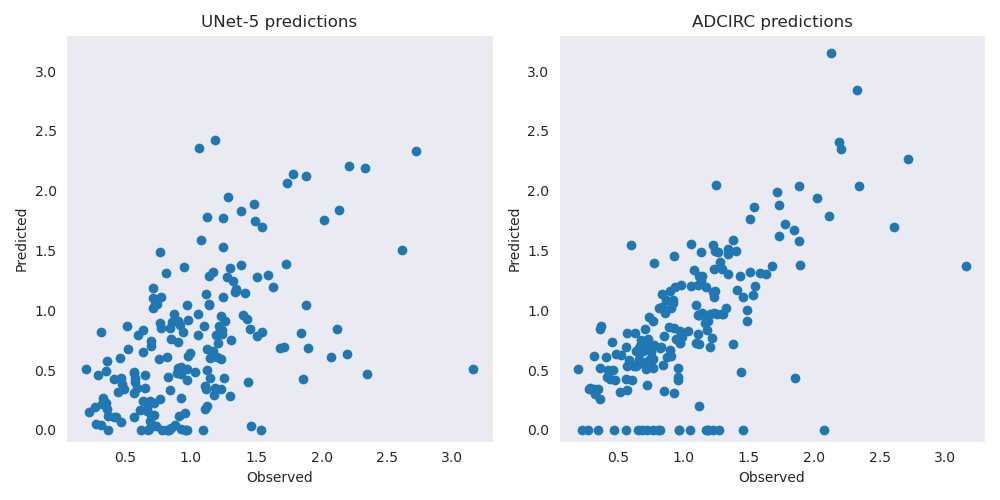}
    \caption{Scatterplot of observed vs predicted maximum water levels for historical hurricanes in the North Atlantic.}
    \label{fig:observation_scatterplot}
\end{figure}

To assess our model's performance on real hurricanes, we utilized a dataset containing track information, meteorological forcing data, and ADCIRC outputs for 67 hurricanes in the North Atlantic basin over a span of 20 years from 2003 to 2023 \cite{cera2023storms}. Observational data from NOAA tidal gauges was available for 30 storms. The majority had data from only a few gauges, but a few made landfall in regions with a higher density of gauges, such as the Texas coastline, and had observational data from 10 or more gauges. For each tidal gauge, we recorded the maximum observed water level, and predictions from both the global UNet-5 model and ADCIRC. There are a total of 188 data points across all 30 storms with data. Figure \ref{fig:observation_scatterplot} contains scatterplot of the predictions from both models compared to the observations. The ADCIRC predictions have a visually better fit to the observational data. This is reflected in the overall RMSE of 0.22 meters for ADCIRC compared to 0.37 meters for the ML model. However, when weighting the points so that each storm has an equal contribution, the RMSE is 0.48 meters for ADCIRC compared to 0.52 meters for UNet-5. The observational data is available on DesignSafe, as are the model predictions, formatted input data, and saved model parameters \cite{designsafe-ml}. It contains the tidal gauge ID, date, and storm name for each observation, as well as predictions from both ADCIRC and UNet-5.

\begin{figure}[H]
    \centering
    \includegraphics[width=1.0\linewidth]{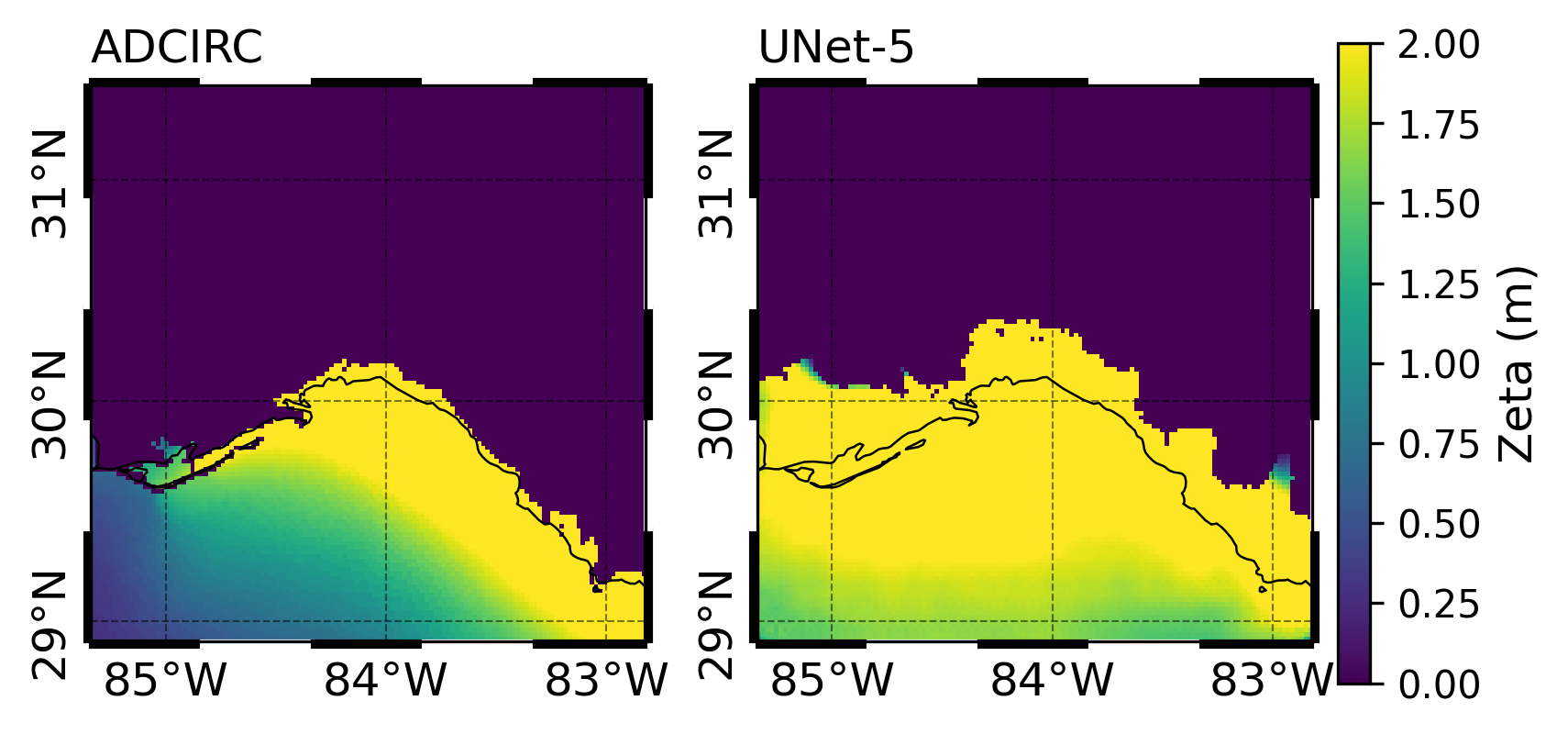}
    \caption{Plot of UNet-5 and ADCIRC peak surge predictions for Hurricane Hermine (2016).}
    \label{fig:hermine_plot}
\end{figure}

For illustrative purposes, we provide a comparison plot of the machine learning model prediction to an ADCIRC hindcast for Hurricane Hermine (2016) in Figure \ref{fig:hermine_plot}.

\section{Concluding Remarks}
\label{sec:conclusion}

Using a novel storm-packing approach, we have generated a global database of ADCIRC simulations of unprecedented size. Due to its scale and coverage, this dataset will help accelerate future surrogate modeling studies. It also has the potential to support efforts unrelated to surrogate modeling, such as flooding risk maps. The comparatively larger scale of the dataset also alleviates the need for imputation strategies and dimensionality reduction developed to handle sparse, small datasets. Finally, the global nature of the data enables training globally applicable models, overcoming the regional limitations imposed by existing databases.

A second contribution is the local formulation of our data-driven model. This solves the high-dimensional output problem without employing dimensionality reduction techniques. It enables direct application of the model to locations not seen in the training data. The model itself attains good prediction accuracy across the globe, and is validated against actual observational data.

The third result is that inclusion of training data from a geographically unrelated area improves the performance of a regional model. This is significant because it is direct evidence of the model's ability to transfer learning between geographic regions. This is a milestone in storm surge emulation, and is a compelling argument to abandon traditional region-specific approaches.

One direction of future work is incorporating physical constraints into our model. This is complicated in the case of maximum surge prediction, because there are no physical equations which directly govern maximum surge profiles. Consequently, such future work would necessitate incorporation of a time-dependent element into the training data. A related effort would be to better incorporate tidal effects into our model, which currently primarily accounts for meteorological and bathymetric effects.

Another important future effort is operationalizing the surrogate model to generate real-time forecasts of peak storm surge for tropical cyclones. The low cost of evaluating the surrogate model would enable making predictions for many potential storm tracks and generating a probabilistic forecast.

\section*{Acknowledgments}
The authors would like to gratefully acknowledge the use of the “DMS23001” and “DMS21031” allocations on the Frontera supercomputer, as well as the "ADCIRC" allocation on the Vista supercomputer, all of which are at the Texas Advanced Computing Center at the University of Texas at Austin.
Author EV  would also like to acknowledge the support of the J.T. Oden Faculty Fellowship Research Program at The Oden Institute for Computational Engineering \& Sciences. 

\section*{CRediT authorship contribution statement}


BP: Writing – original draft, Software., Project administration, Data curation, Conceptualization, Validation,Methodology PA: Writing – original draft, Software, Data curation, Conceptualization, Validation, Methodology, Visualization.  JZ: Writing – original draft, Software, Data curation, Conceptualization, Validation, Methodology.  EV: Writing – original draft, Supervision, Funding acquisition, Resources, Conceptualization, Methodology 

\appendix
\section{Supplementary Figures}\label{secA1}

\begin{figure}[H]
    \centering
    \includegraphics[width=0.8\linewidth]{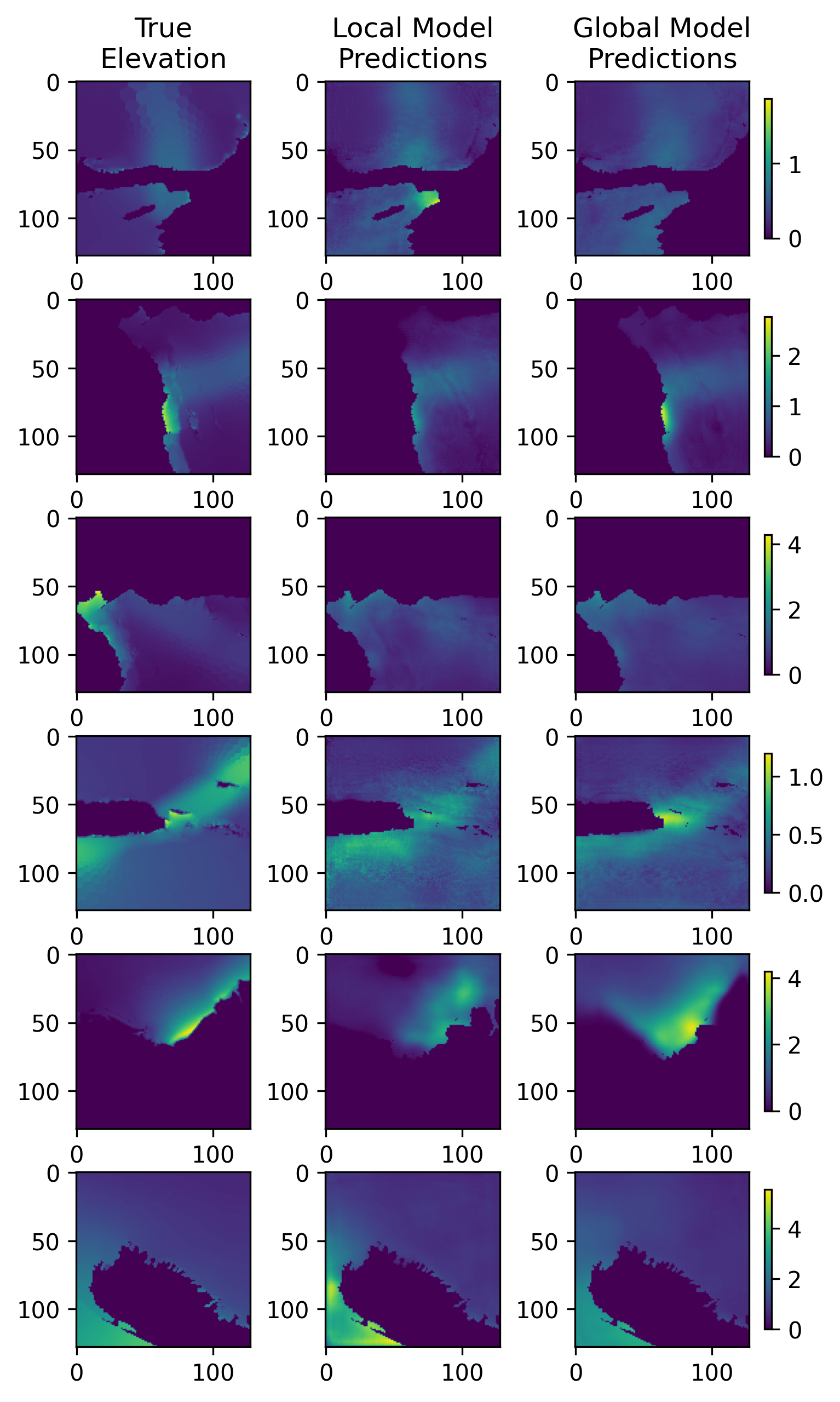}
    \caption{Local vs Global UNet-5 Predictions in the North Atlantic ocean basin.}
    \label{fig:NA}
\end{figure}

\begin{figure}[H]
    \centering
    \includegraphics[width=0.8\linewidth]{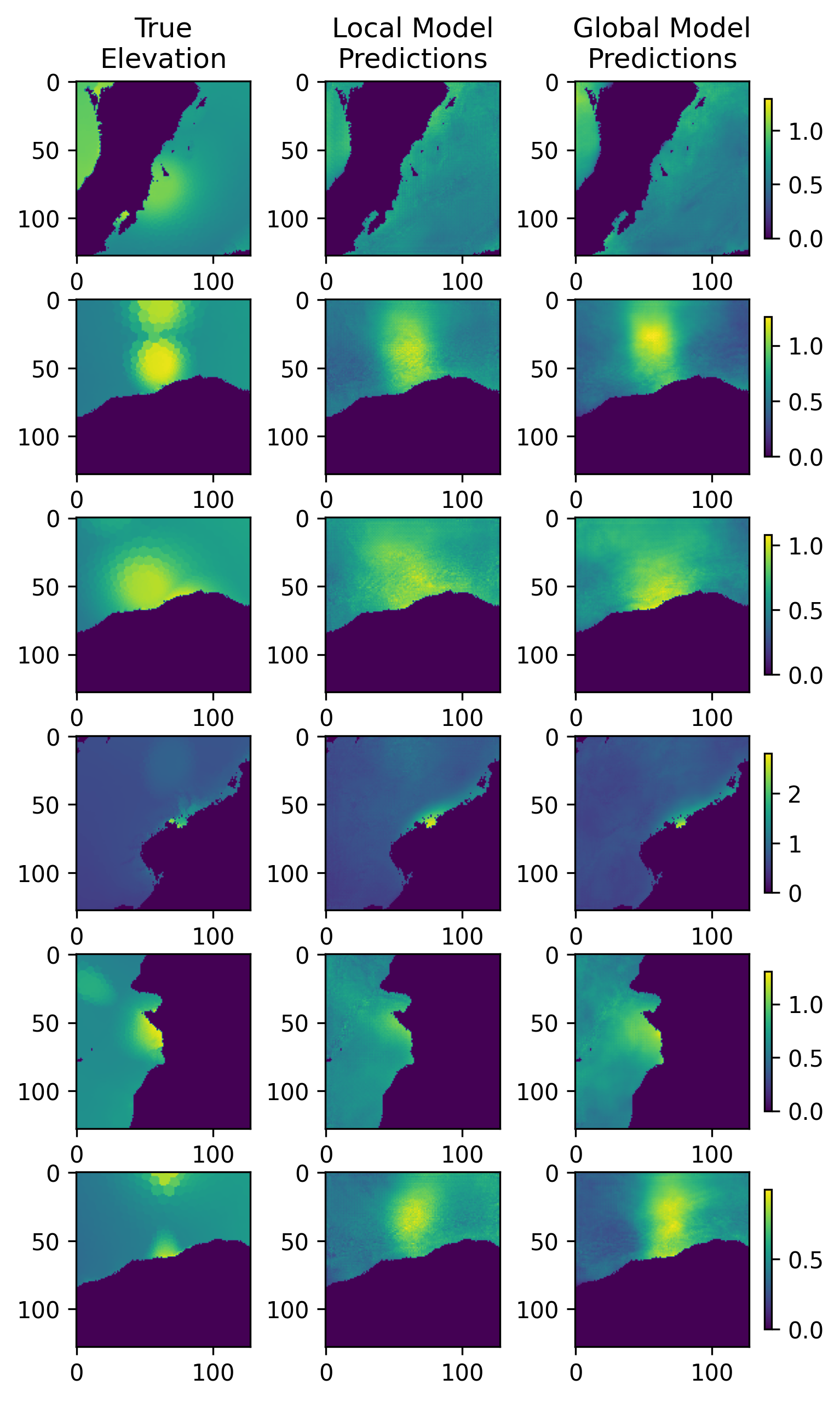}
    \caption{Local vs Global UNet-5 Predictions in the East Pacific ocean basin.}
    \label{fig:EP}
\end{figure}

\begin{figure}[H]
    \centering
    \includegraphics[width=0.8\linewidth]{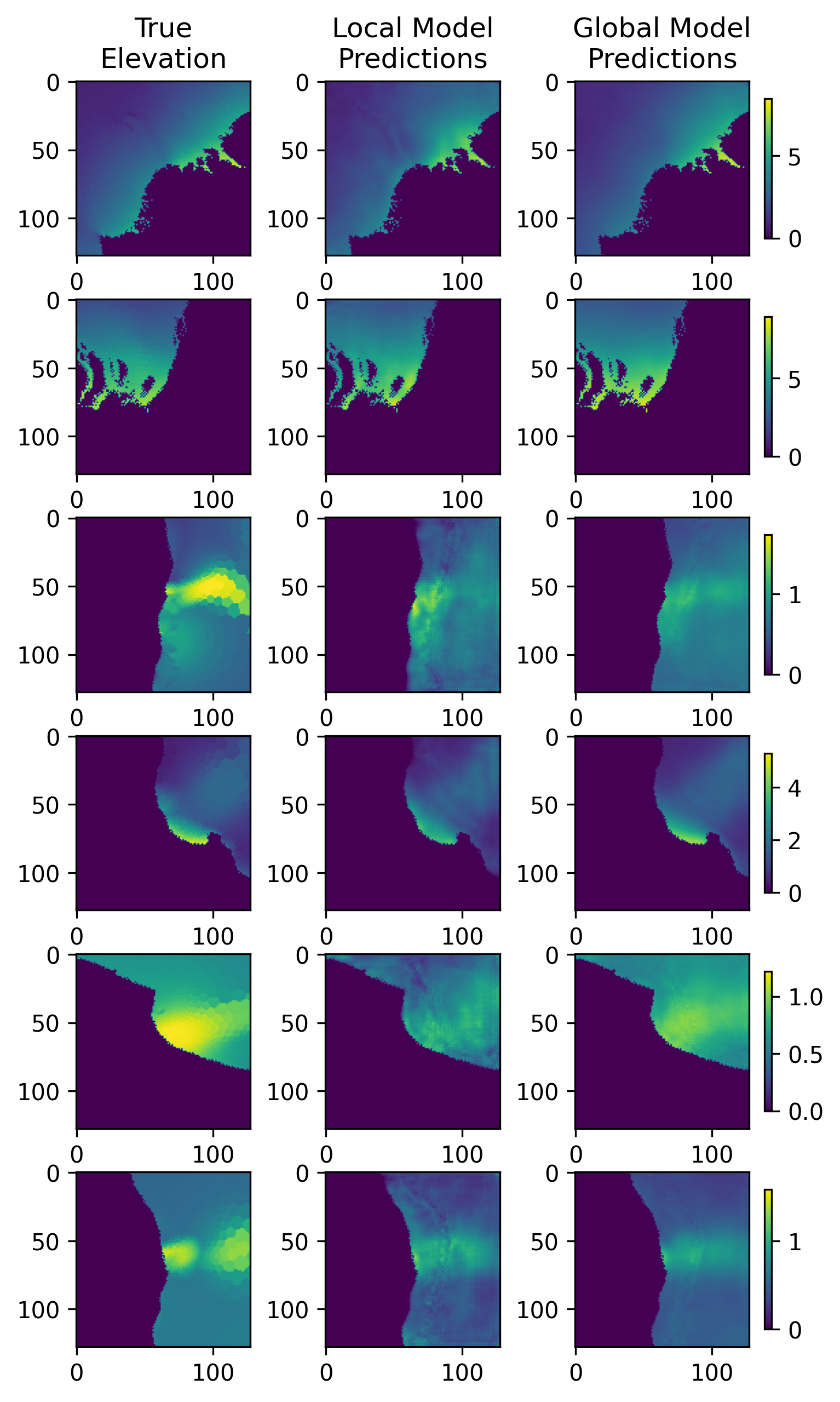}
    \caption{Local vs Global UNet-5 Predictions in the North Indian ocean basin.}
    \label{fig:NI}
\end{figure}

\begin{figure}[H]
    \centering
    \includegraphics[width=0.8\linewidth]{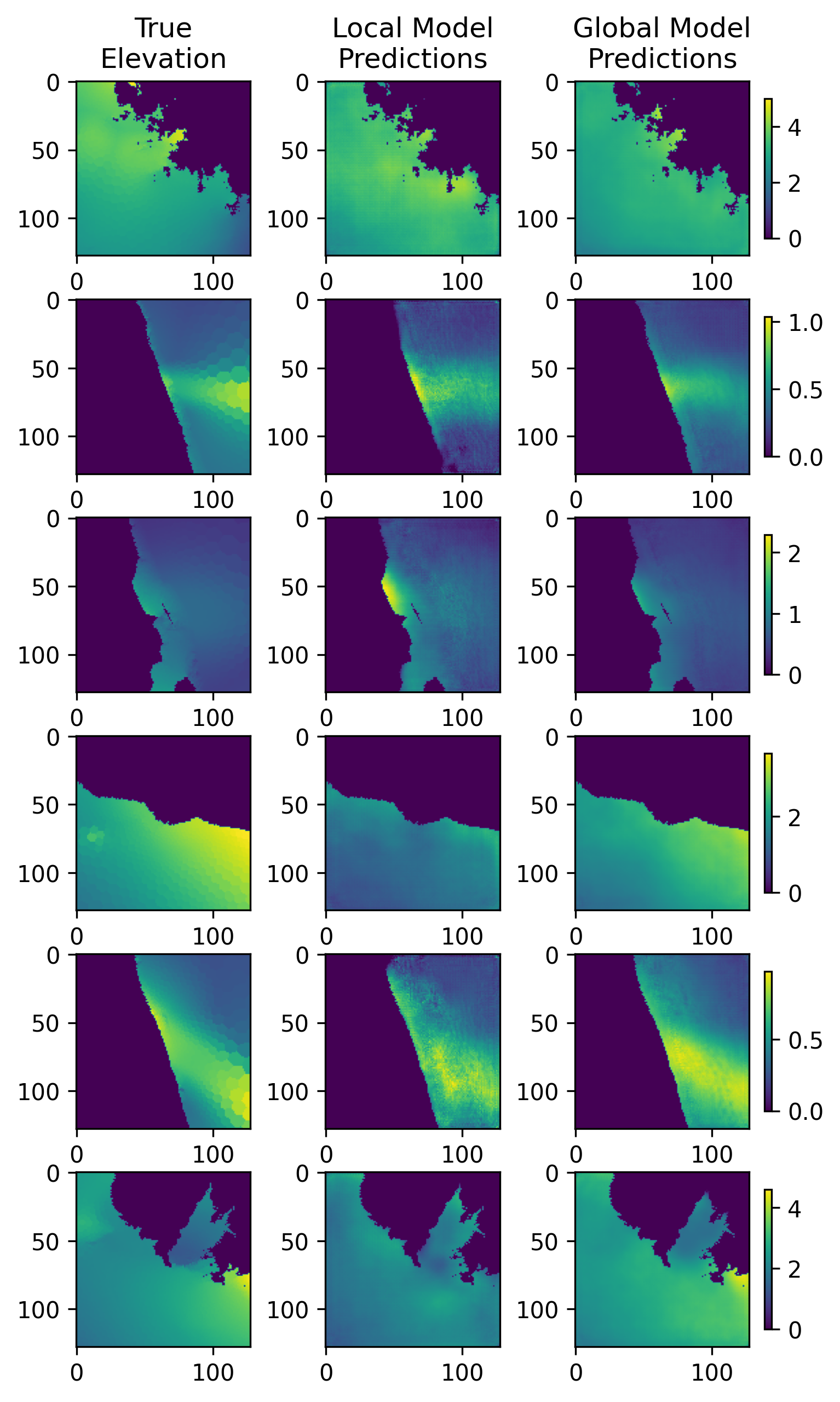}
    \caption{Local vs Global UNet-5 Predictions in the South Indian ocean basin.}
    \label{fig:SI}
\end{figure}

\begin{figure}[H]
    \centering
    \includegraphics[width=0.8\linewidth]{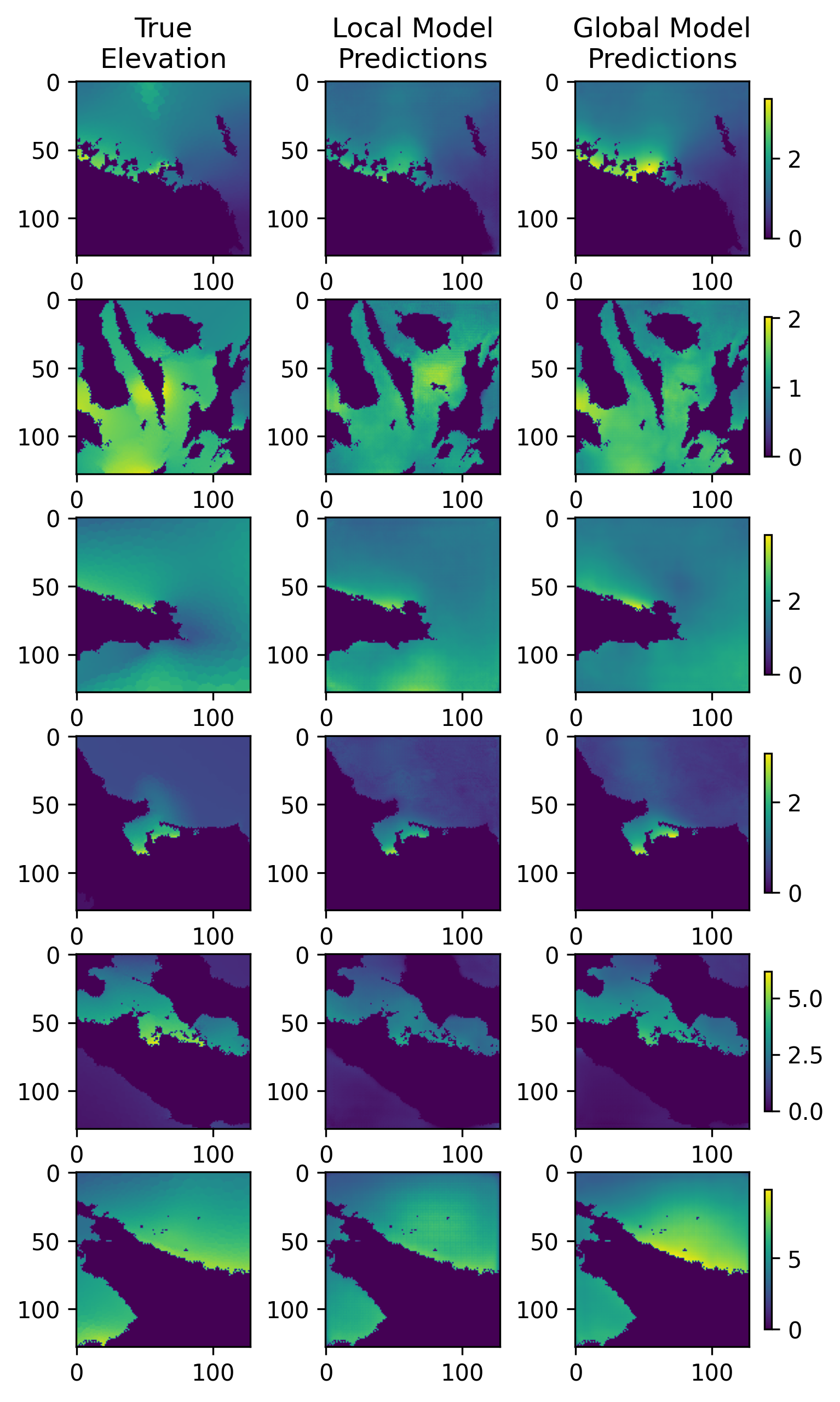}
    \caption{Local vs Global UNet-5 Predictions in the West Pacific ocean basin.}
    \label{fig:WP}
\end{figure}

\begin{figure}[H]
    \centering
    \includegraphics[width=0.8\linewidth]{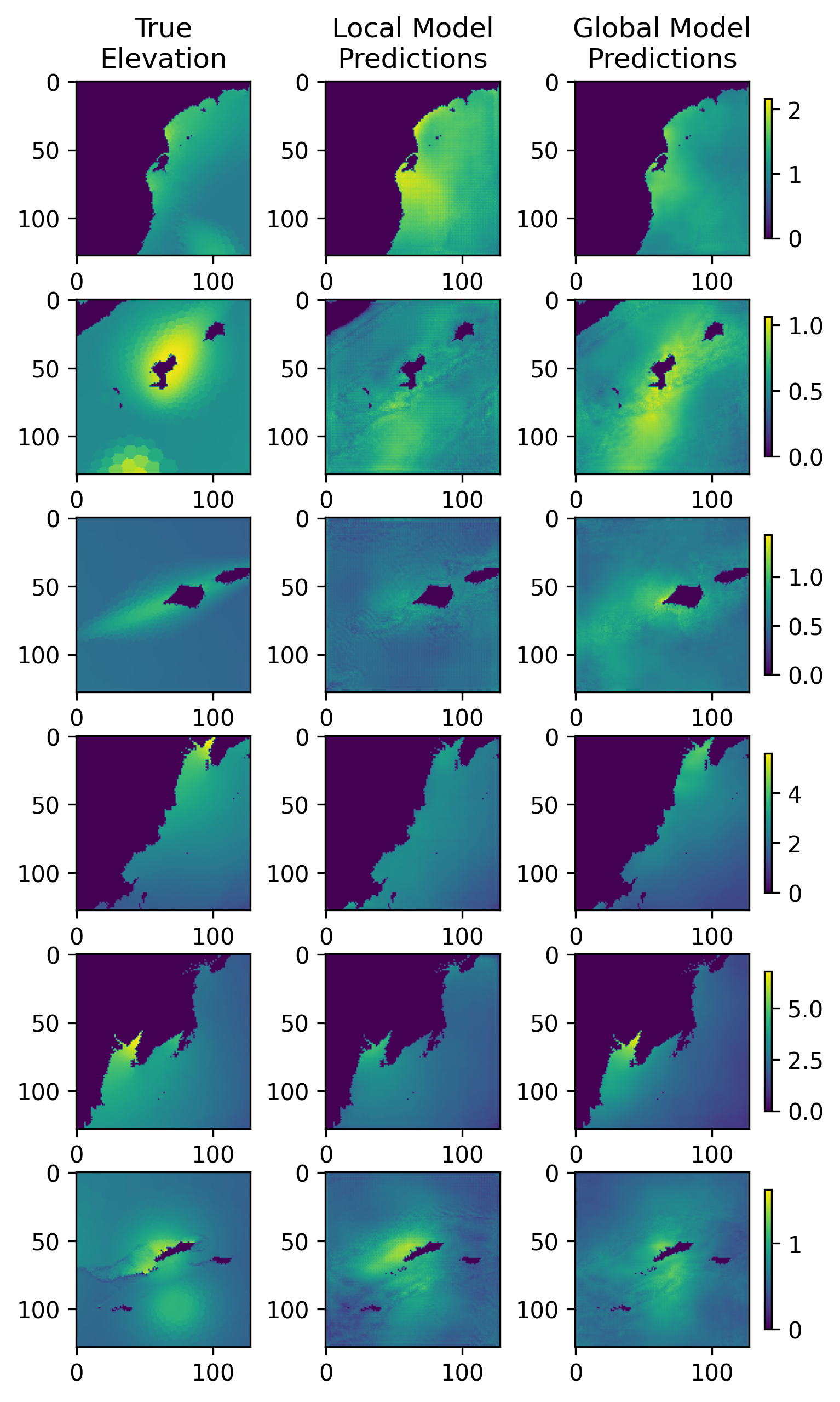}
    \caption{Local vs Global UNet-5 Predictions in the South Pacific ocean basin.}
    \label{fig:SP}
\end{figure}

\bibliographystyle{elsarticle-num}
\bibliography{main}

\end{document}